\documentclass[a4paper,11pt]{article}
\usepackage{pos}
\usepackage{slashed}
\usepackage{graphicx}
\usepackage{epsfig}
\usepackage{amsmath}
\usepackage{color}
\usepackage{float}
\usepackage[rflt]{floatflt}
\usepackage{slashed}
\usepackage{subfig}

\newcommand{\ds}{\displaystyle}
\newcommand{\Mvec}{{\sf M}}
\newcommand{\HA}{{\rm H}}

\newcommand{\gsim}{\raisebox{-0.07cm}{$\, \stackrel{>}{{\scriptstyle
\sim}}\, $}}

\newcommand*\pFqskip{8mu}
\catcode`,\active
\newcommand*\pFq{\begingroup
        \catcode`\,\active
        \def ,{\mskip\pFqskip\relax}%
        \dopFq
}
\catcode`\,12
\def\dopFq#1#2#3#4#5{%
        {}_{#1}F_{#2}\biggl[\genfrac..{0pt}{}{#3}{#4};#5\biggr]%
        \endgroup
}

\title{{\footnotesize DESY-23-089, DO-TH 23/09, CERN-TH-2023-122,
ZU-TH 29/23,  RISC Report Series 23-09, MSUHEP-23-018}\\
Recent 3-Loop Heavy Flavor Corrections to Deep-Inelastic Scattering}
\ShortTitle{Heavy Flavor Corrections to DIS}

\author[a]{J.~Ablinger}
\author[b,c]{A.~Behring}
\author[b]{J.~Bl\"umlein}
\author[b,a]{A.~De Freitas}
\author[b,a]{A.~Goedicke}
\author[d,e]{A.~von~Manteuffel}
\author[a]{C.~Schneider}
\author*[b,f]{K.~Sch\"onwald}

\affiliation[a]{Johannes Kepler University Linz, Research Institute for Symbolic Computation (RISC),
Altenbergerstraße 69, A–4040, Linz, Austria}

\affiliation[b]{Deutsches-Elektronen Synchrotron DESY, Platanenallee 6, 15738 Zeuthen, Germany}

\affiliation[c]{Theoretical Physics Department, CERN, 1211 Geneva 23, Switzerland}

\affiliation[d]{Institut für Theoretische Physik, Universität Regensburg, D-93040 Regensburg, Germany}

\affiliation[e]{Department of Physics and Astronomy, Michigan State University, East Lansing, MI 48824,USA}

\affiliation[f]{Physik-Institut, Universität Zürich, Winterthurerstrasse 190, CH-8057 Zürich, Switzerland}

\emailAdd{jablinge@risc.uni-linz.ac.at}
\emailAdd{Arnd.Behring@cern.ch}
\emailAdd{Johannes.Bluemlein@desy.de}
\emailAdd{abilio.de.freitas@desy.de}
\emailAdd{alexander.goedicke@posteo.de}
\emailAdd{vmante@msu.edu}
\emailAdd{Carsten.Schneider@risc.jku.at}
\emailAdd{kay.schoenwald@physik.uzh.ch}

\abstract{We report on recent progress in calculating the three loop QCD corrections of the heavy flavor 
contributions in deep--inelastic scattering and the massive operator matrix elements of the variable flavor 
number scheme. Notably we deal with the operator matrix elements $A_{gg,Q}^{(3)}$ and $A_{Qg}^{(3)}$ and 
technical steps to their calculation. In particular, a new method to obtain the inverse Mellin transform 
without computing the corresponding $N$--space expressions is discussed.}

\FullConference{
RADCOR2023\\
28th May - 2nd June, 2023\\
Crieff, Scotland, UK\\}

\begin{document}
\maketitle

\section{Introduction}
\label{sec:1}

\vspace*{1mm}
\noindent
Since the last Loops and Legs conference in April 2022, \cite{Blumlein:2022kqz,Blumlein:2021hbq}, 
a series of new results on two-- and three--loop heavy flavor contributions to deep--inelastic scattering 
has been obtained. They concern
the completion of the two--loop corrections to the polarized structure function  $g_1(x,Q^2)$ and the 
two--mass polarized variable flavor number scheme \cite{Bierenbaum:2022biv}, the three--loop unpolarized 
and polarized single--mass operator matrix elements (OMEs) $(\Delta) A_{gg,Q}^{(3)}$ 
\cite{Ablinger:2022wbb}, and first analytic steps 
beyond first--order factorizable contributions to $A_{Qg}^{(3)}$ in \cite{Blumlein:2017wxd}, given 
in Ref.~\cite{Behring:2023rlq}. Herewith, only the constant part in the dimensional parameter $\varepsilon  
= D-4$ to the unrenormalized OMEs $(\Delta) A_{Qg}^{(3)}$ have still to be completed as a function of 
general values of $N$, to obtain the full description of the heavy flavor contributions to deep--inelastic 
scattering at three--loop order at large enough virtualities. The known contributions have recently been 
discussed in  Ref.~\cite{Blumlein:2023aso}.
In this report we will concentrate on the results obtained in Refs.~\cite{Ablinger:2022wbb,
Behring:2023rlq}.
\section{\boldmath The massive OMEs $A_{gg,Q}^{(3)}$ and $\Delta A_{gg,Q}^{(3)}$}
\label{sec:2}

\vspace*{1mm}
\noindent
These three--loop OMEs form an essential asset to the description of the three--loop variable flavor 
number scheme, allowing heavy quarks to become light in the asymptotic region $Q^2 \gg m^2$. With this 
all OMEs, except $(\Delta) A_{Qg}^{(3)}$, are known completely. A preliminary closed form $N$-space result 
for all 
even 
moments has been derived by us in 2015 \cite{Ablinger:2015lvm}. However, there has been a problem with the 
analytic 
continuation to $x$--space with one Feynman diagram. We finally could obtain this by applying the methods 
of 
Ref.~\cite{Behring:2023rlq} in Ref.~\cite{Ablinger:2022wbb}. The final expression could then be transformed
to $N$--space again, which is of advantage for $N$--space evolution codes.

The calculation of the gluonic OMEs proceeded by using a series of computation techniques, described in 
Refs.~[63--73] of \cite{Ablinger:2022wbb}.  The problem at hand is finally characterized by 
the alphabet 
\begin{eqnarray}
\mathfrak{A}_1 = \Biggl\{\frac{1}{x}, \frac{1}{1-x}, \frac{1}{1+x},
\frac{\sqrt{1-x}}{x}, \sqrt{x(1-x)}, \frac{1}{\sqrt{1-x}} \Biggr\}.
\end{eqnarray}
In $N$--space also different finite (inverse) binomial sums are contributing, an example of which is
\begin{eqnarray}
{\sf BS}_9(N) &=&
\sum_{\tau_1=1}^N \frac{\ds 4^{-\tau_1} \big(
        2 \tau_1\big)!
\sum_{\tau_2=1}^{\tau_1} \frac{\ds 4^{\tau_2} \big(
        \tau_2!\big)^2
\sum_{\tau_3=1}^{\tau_2} \ds \frac{1}{\tau_3}}{\big(
        2 \tau_2\big)! \tau_2^2}}{\big(
        \tau_1!\big)^2 \tau_1}.
\end{eqnarray}
These binomial sums obey recursively first--order difference equations and their asymptotic expansions
can be computed analytically in their analyticity region. Furthermore, one may calculate their Mellin 
inversion to $x$--space analytically and obtain iterated integrals over $\mathfrak{A}_1$ there.
The small and large $x$ expansions can be obtained analytically. 

\begin{figure}
\centering
\includegraphics[width=0.7\textwidth]{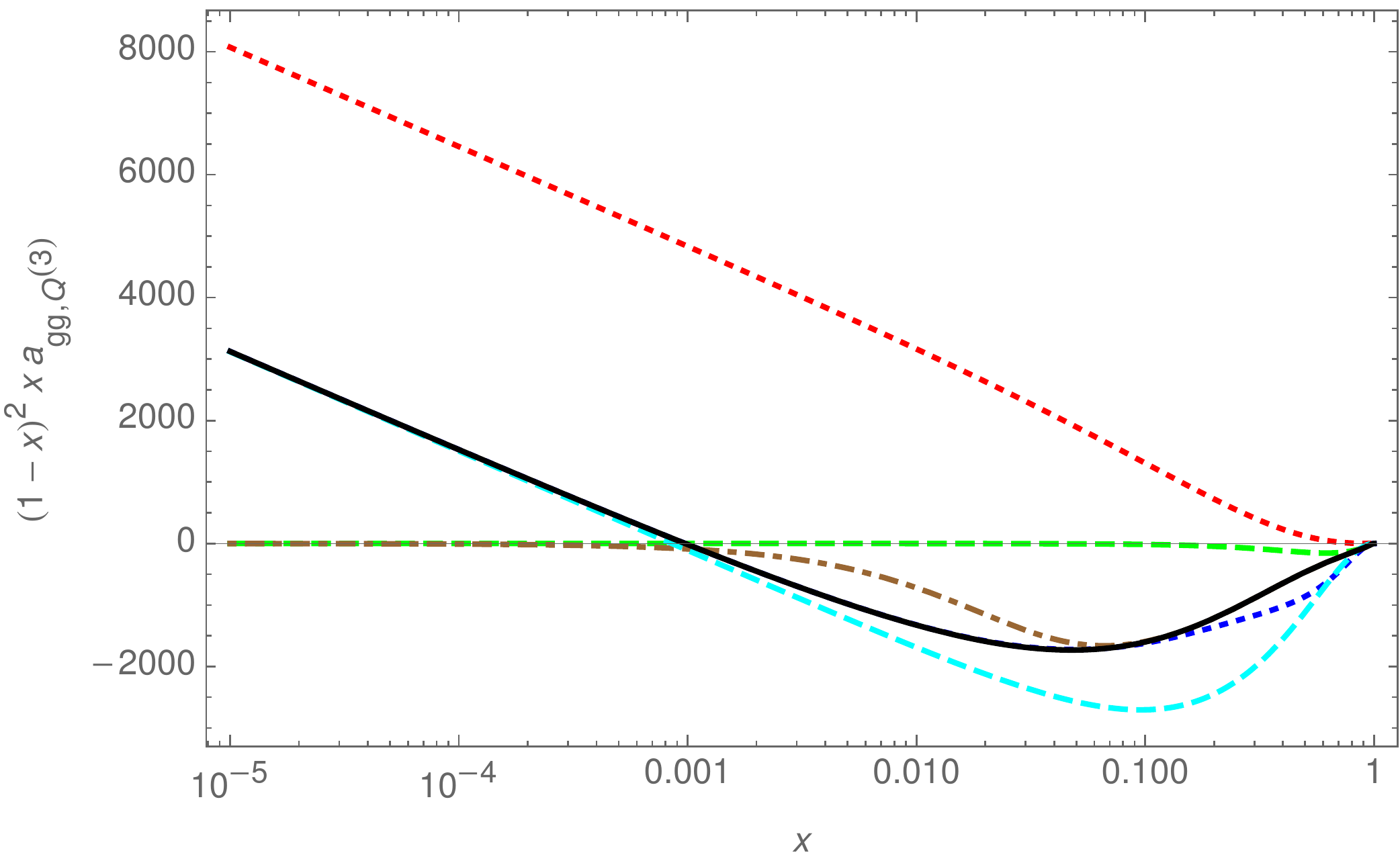}
\caption{
The non--$N_F$ terms of $a_{gg,Q}^{(3)}(N)$ (rescaled) as a function of $x$.
Solid line (black): complete result;
upper dotted line (red): term $\propto \ln(x)/x$;
lower dashed line (cyan): small $x$ terms $\propto 1/x$;
lower dotted line (blue): small $x$ terms including all $\ln(x)$ terms
up the constant term;
upper dashed line (green): large $x$ contribution
up to the constant term;
dash-dotted line (brown): full large $x$ contribution; from Ref.~\cite{Ablinger:2022wbb}.}
\label{fig:1}
\end{figure}

Numerically, 50-term expansions around $x=0$ and $x=1$ are sufficient to obtain the constant parts of the 
unrenormalized OMEs $(\Delta) A_{gg,Q}^{(3)}$ to high precision, cf. Figure~\ref{fig:1}. The so--called 
BFKL limit thoroughly deviates from the complete results, which is a well--known fact observed in many 
phenomenological cases \cite{Blumlein:1997em}. Several sub--leading terms are necessary to yield a 
quantitative description of the small $x$ region. It is also visible that the expansions around 
$x=0$ and $x=1$ are sufficient to describe the full function as mentioned above.
\section{Inverse Mellin transform from a resummed variable}
\label{sec:3}

\vspace*{1mm}
\noindent
Already in Ref.~\cite{Ablinger:2022wbb} it has been necessary to work
in $x$--space in part. For this reason the resummation
\begin{eqnarray}
\sum_{N = 1}^\infty t^N (\Delta.p)^{N-1} = \frac{t}{1 - t \Delta.p} 
\end{eqnarray}
has been performed \cite{Ablinger:2014yaa}. This way, the original discrete Mellin variable $N$ is 
transformed into the continuous variable 
$t$. The respective moments are obtained by a formal Taylor expansion. In Ref.~\cite{Behring:2023rlq} we 
have shown how to obtain from the resummed form the associated inverse Mellin transform in $x$--space.
Since some of the contributing terms are distribution--valued, one should deal with them first. They can 
be structurally identified in $t$--space as those leading to the distributions $\propto \delta(1-x)$ and 
$[\ln^k(1-x)/(1-x)]_+,~~k \in \mathbb{N}, k \geq 0$. The remaining $t$--space expressions will then lead 
to regular contributions in $x$--space, if the complete amplitude is considered.
The method of Ref.~\cite{Behring:2023rlq} works for contributions 
which obey both first--order factorizing and non first-order factorizing differential
  equations.
In the former case the known classes are harmonic polylogarithms, 
generalized harmonic polylogarithms, cyclotomic harmonic polylogarithms, and iterated integrals also 
containing square--root valued letters. Non first--order factorizing cases are those starting with 
$_2F_1$--solutions, cf. e.g. \cite{Ablinger:2017bjx}, and generalizations thereof, obeying even higher 
order differential equations.

Let us consider an example in the first--order factorizing case, e.g. the regular function
\begin{eqnarray}
\tilde{F}_1(t)  = \HA_{0,0,1}(t),
\end{eqnarray}
with $\HA_{\vec{a}}(t)$ denoting a harmonic polylogarithm. 
One obtains 
\begin{eqnarray}
\label{eq:a}
F_1(x)  = \frac{1}{\pi }{\sf Im} \tilde{F}_1\left(\frac{1}{x}\right)  = \frac{1}{2} \HA^2_0(x).
\end{eqnarray}
Here one has to consider the monodromy around $t=1$ only, while in general this is necessary for 
$t = \pm 1$. The Mellin transform of Eq.~(\ref{eq:a}) is 
\begin{eqnarray}
\Mvec[F_1(x)](N) = \int_0^1 dx x^{N-1} F_1(x)  = \frac{1}{N^3}.
\end{eqnarray}
In accordance to this one obtains the formal Taylor series
\begin{eqnarray}
\tilde{F}_1(t) = \sum_{k=1}^\infty \frac{t^k}{k^3}.
\end{eqnarray}

Let us now consider the case of non first--order factorizing differential equations. Here one has to deal 
with a higher linear system to be decoupled, where at least a $2 \times 2$ system remains. As has been 
shown in Ref.~\cite{Behring:2023rlq}, it is important which of the solutions is dealt with first, since
the $\varepsilon$--expansion of the solution in $x$--space may have a simpler structure than the case for 
other choices.

The alphabets over which the iterated integrals in $t$--space are obtained do now contain also factors
of higher transcendental functions and derivatives thereof, unlike for the first--order factorizing cases.
E.g. one has
\begin{eqnarray}
\mathfrak{A}_2 = \Biggl\{
\frac{1}{t}, \frac{1}{1-t}, \frac{1}{8+t}, g_1(t), g_2(t), \frac{g_1(t)}{t}, ....
\Biggr\},
\end{eqnarray}
where
\begin{eqnarray}
\label{eq:HEUN3a}
g_1(t) &=&
\frac{2}{(1-t)^{2/3}(8+t)^{1/3}}
\pFq{2}{1}{\frac{1}{3},\frac{4}{3}}{2}{-\frac{27 t}{(1-t)^2(8+t)}},
\\
\label{eq:HEUN3b}
g_2(t) &=& \frac{2}{(1-t)^{2/3}(8+t)^{1/3}}
\pFq{2}{1}{\frac{1}{3},\frac{4}{3}}{\frac{2}{3}}{1+\frac{27 t}{(1-t)^2(8+t)}}.
\end{eqnarray}
The closed form solutions now allow to form the corresponding iterative integrals over $\mathfrak{A}_2$ 
and
the analytic continuation of $t \rightarrow \pm 1/x$. In part, regularizations are necessary. Furthermore,
iterative integrals at $x=1$ will occur as constants in the description of the problem. Not all of them 
reduce to known special numbers and they have to be computed numerically to high precision in the end.
The final expressions obtained can now be expanded around the points $x = 0, 1$ and e.g. also 1/2 to obtain
fast and highly precise numerical representations in the region $x \in [0,1]$, 
cf.~Ref.~\cite{Behring:2023rlq}, which is necessary 
for phenomenological applications.

\section{Conclusions}
\label{sec:4}

\vspace*{1mm}
\noindent
The calculation of the unpolarized and polarized single-- and two--mass three--loop contributions 
to deep--inelastic scattering started with a series of moments in 2009 \cite{Bierenbaum:2009mv} and 
then turned
to the general $N$ and $x$--space results in the limit $Q^2 \gg m_Q^2$ thereafter. 
In the region $Q^2/m_Q^2 \gsim 10$ this is sufficient for considering the structure function $F_2(x,Q^2)$,
cf.~\cite{Buza:1995ie}. A cut of this kind should be applied because of higher twist contributions in the 
data at lower scales.

Along with these calculations various mathematical and 
computer--algebraic technologies were developed, cf. Refs.~\cite{Blumlein:2018cms,Blumlein:2022qci} for 
surveys. 
After a very successful treatment of the majority of problems in $N$ space by using the method of arbitrary 
high Mellin moments \cite{Blumlein:2017dxp}, the method of guessing \cite{Blumlein:2009tj}, 
 and difference field theory 
as implemented in the package 
{\tt Sigma} \cite{SIG1,SIG2}, we had to refer to $x$--space in different cases 
to obtain the complete result. Here the method 
described in Ref.~\cite{Behring:2023rlq} has been used extensively. For a brief report on the 
physical status reached for the heavy flavor corrections see \cite{Blumlein:2023aso}. After having obtained 
the OMEs $(\Delta) A_{gg,Q}^{(3)}$, 
current work is dedicated to complete $(\Delta) A_{Qg}^{(3)}$, since now also the massless polarized 
three--loop Wilson coefficients are available \cite{Blumlein:2022gpp} and the unpolarized ones 
\cite{Vermaseren:2005qc} are confirmed. $(\Delta) A_{Qg}^{(3)}$ contains 
$_2F_1$--sectors requiring special attendance. In the internal representation also the different
functional contributions to first--order factorizing terms may show exponential growth in the limit
$N \rightarrow \infty$, which will cancel, however, between the different mathematical classes 
of contributions. One is advised to show this analytically.

\vspace*{2mm}
\noindent
{\bf Acknowledgment.}~This work has received funding in part the European Union’s Horizon 2020 
research and innovation programme grant agreement 101019620 (ERC Advanced Grant TOPUP),
EU TMR network SAGEX agreement No. 764850 (Marie Sklodowska-Curie) the Austrian Science Fund (FWF) 
grants SFB F50 (F5009–N15) and P33530 and by the Research Center “Elementary Forces and Mathematical 
Foundations (EMG)” of J. Gutenberg University Mainz and DFG.


\begin{thebibliography}{99}
%
\bibitem{Blumlein:2022kqz}
J.~Bl\"umlein, P.~Marquard, C.~Schneider and K.~Sch\"onwald,
{\it The 3-loop anomalous dimensions from off-shell operator matrix 
elements},
PoS (LL2022) 048
[arXiv:2207.07943 [hep-ph]].
%
\bibitem{Blumlein:2021hbq}
J.~Bl\"umlein, M.~Saragnese and C.~Schneider,
{\it Hypergeometric Structures in Feynman Integrals},
Annals of Mathematics and Artificial Intelligence, (2023) in print,
[arXiv:2111.15501 [math-ph]].
%
\bibitem{Bierenbaum:2022biv} 
I.~Bierenbaum, J.~Bl\"umlein, A.~De Freitas, A.~Goedicke, S.~Klein and K.~Sch\"onwald,
{\it $O(\alpha_s^2)$ polarized heavy flavor corrections to
deep-inelastic scattering at $Q^2 \gg m^2$},
Nucl. Phys. B \textbf{988} (2023) 116114
[arXiv:2211.15337 [hep-ph]].
%
\bibitem{Ablinger:2022wbb}
J.~Ablinger, A.~Behring, J.~Bl\"umlein, A.~De Freitas, A.~Goedicke, A.~von Manteuffel, C.~Schneider 
and K.~Sch\"onwald,
{\it The unpolarized and polarized single-mass three-loop heavy flavor operator matrix elements 
A$_{gg,Q}$ and \ensuremath{\Delta}A$_{gg,Q}$},
JHEP \textbf{12} (2022) 134
[arXiv:2211.05462 [hep-ph]].
%
\bibitem{Blumlein:2017wxd}
J.~Bl\"umlein, J.~Ablinger, A.~Behring, A.~De Freitas, A.~von Manteuffel, C.~Schneider and C.~Schneider,
{\it Heavy Flavor Wilson Coefficients in Deep-Inelastic Scattering: Recent Results},
PoS (QCDEV2017)  031
[arXiv:1711.07957 [hep-ph]].
%
\bibitem{Behring:2023rlq}
A.~Behring, J.~Bl\"umlein and K.~Sch\"onwald,
{\it The inverse Mellin transform via analytic continuation},
JHEP \textbf{06} (2023) 062
[arXiv:2303.05943 [hep-ph]].
%
\bibitem{Ablinger:2015lvm}
J.~Ablinger, A.~Behring, J.~Bl\"umlein, A.~De Freitas, A.~Hasselhuhn, A.~von Manteuffel, C.G.~Raab, 
M.~Round, C.~Schneider and F.~Wi\ss{}brock,
{\it 3-Loop Corrections to the Heavy Flavor Wilson Coefficients in Deep-Inelastic Scattering},
PoS (EPS-HEP2015) 504
[arXiv:1602.00583 [hep-ph]].
%
\bibitem{Blumlein:2023aso}
J.~Bl\"umlein,
{\it Deep-Inelastic Scattering: What do we know ?},
Memorial Volume for Harald Fritzsch, (World Scientific, Singapore) to appear,
[arXiv:2306.01362 [hep-ph]].
%
\bibitem{Blumlein:1997em}
J.~Bl\"umlein and A.~Vogt,
{\it The Evolution of unpolarized singlet structure functions at small x},
Phys. Rev. D \textbf{58} (1998) 014020
[hep-ph/9712546].
%
\bibitem{Ablinger:2014yaa}
J.~Ablinger, J.~Bl\"umlein, C.~Raab, C.~Schneider and F.~Wi\ss{}brock,
{\it Calculating Massive 3-loop Graphs for Operator Matrix Elements by the Method of Hyperlogarithms},
Nucl. Phys. B \textbf{885} (2014) 409--447
[arXiv:1403.1137 [hep-ph]].
%
\bibitem{Ablinger:2017bjx}
J.~Ablinger, J.~Bl\"umlein, A.~De Freitas, M.~van Hoeij, E.~Imamoglu, C.~G.~Raab, C.~S.~Radu and C.~Schneider,
{\it Iterated Elliptic and Hypergeometric Integrals for Feynman Diagrams},
J. Math. Phys. \textbf{59} (2018) no.6, 062305
[arXiv:1706.01299 [hep-th]].
%
\bibitem{Bierenbaum:2009mv}
I.~Bierenbaum, J.~Bl\"umlein and S.~Klein,
{\it Mellin Moments of the $O(\alpha^3_s)$ Heavy Flavor Contributions to unpolarized Deep-Inelastic 
Scattering at $Q^2 \gg m^2$ and Anomalous Dimensions},
Nucl. Phys. B \textbf{820} (2009) 417--482
[arXiv:0904.3563 [hep-ph]].
%
\bibitem{Buza:1995ie}
M.~Buza, Y.~Matiounine, J.~Smith, R.~Migneron and W.L.~van Neerven,
{\it Heavy quark coefficient functions at asymptotic values $Q^2 \gg m^2$}
Nucl. Phys. B \textbf{472} (1996) 611--658
[hep-ph/9601302].
%
\bibitem{Blumlein:2018cms}
J.~Bl\"umlein and C.~Schneider,
{\it Analytic computing methods for precision calculations in quantum field theory},
Int. J. Mod. Phys. A \textbf{33} (2018) no.17, 1830015
[arXiv:1809.02889 [hep-ph]].
%
\bibitem{Blumlein:2022qci}
J.~Bl\"umlein and C.~Schneider,
{\it The SAGEX review on scattering amplitudes Chapter 4: Multi-loop Feynman integrals},
J. Phys. A \textbf{55} (2022) no.44, 443005
[arXiv:2203.13015 [hep-th]].
%
\bibitem{Blumlein:2017dxp}
J.~Bl\"umlein and C.~Schneider,
{\it The Method of Arbitrarily Large Moments to Calculate Single Scale Processes in Quantum Field Theory},
Phys. Lett. B \textbf{771} (2017) 31--36
[arXiv:1701.04614 [hep-ph]].
%
\bibitem{Blumlein:2009tj}
J.~Bl\"umlein, M.~Kauers, S.~Klein and C.~Schneider,
{\it Determining the closed forms of the $O(a_s^3)$ anomalous dimensions and Wilson coefficients from 
Mellin moments by means of computer algebra},
Comput. Phys. Commun. \textbf{180} (2009) 2143--2165
[arXiv:0902.4091 [hep-ph]].
%
\bibitem{SIG1}
C.~Schneider, {\it Symbolic Summation Assists Combinatorics},
{S\'em.~Lothar. Combin.\/} {\bf 56} (2007) 1--36
 article B56b.
%
\bibitem{SIG2}
C.~Schneider, {\it Simplifying Multiple Sums in Difference Fields}, in:~{{\sf Computer
Algebra in Quantum Field Theory: Integration,
  Summation and Special Functions}\/} Texts and Monographs in Symbolic
  Computation eds. C.~Schneider and J.~Bl\"umlein  (Springer, Wien, 2013) 325--360
  [arXiv:1304.4134 [cs.SC]].
%
\bibitem{Blumlein:2022gpp}
J.~Bl\"umlein, P.~Marquard, C.~Schneider and K.~Sch\"onwald,
{\it The massless three-loop Wilson coefficients for the deep-inelastic structure functions F$_{2}$, 
F$_{L}$, xF$_{3}$ and g$_{1}$},
JHEP \textbf{11} (2022) 156
[arXiv:2208.14325 [hep-ph]].
%
\bibitem{Vermaseren:2005qc}
J.A.M.~Vermaseren, A.~Vogt and S.~Moch,
{\it The Third-order QCD corrections to deep-inelastic scattering by photon exchange},
Nucl. Phys. B \textbf{724} (2005) 3--182
[hep-ph/0504242].
\end{thebibliography}
\end{document}